\documentclass{Rinton-P9x6}

\def\lapproxeq{\lower .7ex\hbox{$\;\stackrel{\textstyle
<}{\sim}\;$}}
\def\gapproxeq{\lower .7ex\hbox{$\;\stackrel{\textstyle
>}{\sim}\;$}}

\begin{document}

\title{
F-term Inflation and Leptogenesis with 5D $SO(10)$\footnote{Based on 
the work collaborated with Q. Shafi (Phys. Lett. {\bf B556}, 97 (2003) 
[hep-ph/0211059])
}  
}

\author{
Bumseok Kyae
}
\address{
Bartol Research Institute, University of Delaware, Newark, DE 19716, USA\\
E-mail: bkyae@bartol.udel.edu
}

\maketitle

\abstracts{
We discuss a five dimensional inflationary scenario based on
a supersymmetric $SO(10)$ model compactified on $S^1/(Z_2\times Z_2')$.   
Inflation is implemented through scalar potentials    
on four dimensional branes, and a brane-localized Einstein-Hilbert term 
is essential to make both brane vacuum energies positive 
during inflation.        
The orbifold boundary conditions break the $SO(10)$ gauge symmetry 
to $SU(4)_c\times SU(2)_L\times SU(2)_R$ ($\equiv H$).     
The inflationary scenario yields $\delta T/T\propto (M/M_{\rm Planck})^2$, 
which fixes $M$, the symmetry breaking scale of $H$ to be close 
to the SUSY GUT scale of $10^{16}$ GeV.   
The scalar spectral index $n$ is $0.98-0.99$, 
and the tensor to scalar ratio $r$ is $\lapproxeq 10^{-4}$.  
%
%
The inflaton decay into the lightest right handed neutrinos
yields the observed baryon asymmetry via leptogenesis.
} 

\section{Introduction}

Supersymmetric grand unified theories (SUSY GUTs) provide  
a prominent framework for physics beyond the standard model, and
it is therefore natural to ask if there exists in this framework 
an intimate connection with inflation. 
Indeed, a class of realistic SUSY inflationary models   
elegantly addresses the question~\cite{hybrid}.  
In a promising SUSY inflationary model~\cite{khalil} 
based on the $SO(10)$ subgroup
$SU(4)_c\times SU(2)_L\times SU(2)_R$ ($\equiv H$)~\cite{ps}, for instance,   
the scalar spectral index $n$ in the model has a value very close to unity 
(typically $n\approx 0.98-0.99$) in excellent agreement with a variety of 
observations including the recent WMAP data~\cite{cobe}.    
%
%
In particular, the quadrupole microwave anisotropy is proportional
to $(M/M_ {\rm Planck})^2$,
where $M$ denotes the gauge symmetry breaking scale of $H$, 
and $M_{\rm Planck}=1.2\times 10^{19}$ GeV.
Thus, $M$ is expected to be of order $10^{16}$ GeV, 
which is quite close to the supersymmetric grand unification scale 
inferred from the evolution of
the minimal supersymmetric standard model (MSSM) gauge couplings.  
The vacuum energy density during inflation is of order $10^{14}$ GeV,         
so that the gravitational contribution to the quadrupole anisotropy
is essentially negligible.  
The inflaton field in this scenario eventually decays into right handed
neutrinos, whose out of equilibrium decays lead to 
leptogenesis~\cite{lepto,ls}.
However, a straightforward extension to the full $SO(10)$ model is 
obstructed by the notorious doublet-triplet splitting problem.  

Orbifold symmetry breaking in higher dimensional GUTs
have recently attracted a great deal of attention
because the two particularly pressing problems encountered
in four dimensional (4D) SUSY GUTs, 
namely, the doublet-triplet splitting problem
and the dimension five proton decay problem 
are easily circumvented without fine-tuning of parameters~\cite{kawamura}.  
The existence of the orbifold dimension 
can readily break a grand unified symmetry such as $SO(10)$ 
to its maximal subgroup $H$~\cite{dermisek}.  
%
%
Our objective here is to take advantage of recent orbifold constructions
of five dimensional (5D) supersymmetric $SO(10)$, and 
provide a 5D framework which can be merged 
with the four dimensional (4D) supersymmetric inflationary scenario 
based on $H$~\cite{original}.  
Because of $N=2$ SUSY (in 4D sense) in 5D bulk, the F-term inflaton 
potential is allowed only on the 4D orbifold fixed points (branes), 
where only $N=1$ SUSY is preserved.    
%
%

\section{F-term Inflation}

The four dimensional inflationary model is best illustrated by considering
the following superpotential which allows the breaking of some gauge
symmetry $G$ down to $SU(3)_c\times SU(2)_L\times U(1)_Y$,   
keeping supersymmetry (SUSY) intact \cite{hybrid,lyth}:
\begin{eqnarray} \label{simplepot}
W_{\rm infl}=\kappa S(\phi\bar{\phi}-M^2) ~.
\end{eqnarray}
Here $\phi$ and $\bar{\phi}$ represent superfields
whose scalar components acquire non-zero vacuum expectation values (VEVs).
For the particular example of $G=H$ above,
they belong to the ${\bf (\overline{4}, 1,2)}$ and ${\bf (4,1,2)}$
representations of $H$.
The $\phi$, $\bar{\phi}$ VEVs break $H$ to the MSSM gauge group.  
The singlet superfield $S$ provides the scalar field that drives inflation.
Note that by invoking a suitable $R$ symmetry $U(1)_R$,
the form of $W$ is unique at the renormalizable level, and it 
is gratifying to realize that $R$ symmetries naturally occur 
in (higher dimensional) supersymmetric theories
and can be appropriately exploited.
From $W$, it is straightforward to show that the supersymmetric minimum
corresponds to non-zero (and equal in magnitude) VEVs
for $\phi$ and $\bar{\phi}$, while $\langle S\rangle =0$.
(After SUSY breaking {\it $\grave{a}$ la} $N=1$ supergravity 
(SUGRA), $\langle S\rangle$ acquires a VEV 
of order $m_{3/2}$ (gravitino mass)).

An inflationary scenario is realized in the early universe
with both $\phi$, $\bar{\phi}$ and $S$ displaced
from their present day minima.
Thus, for $S$ values in excess of the symmetry breaking scale $M$,
the fields $\phi$, $\bar{\phi}$ both vanish,
the gauge symmetry is restored, and a potential energy density  
$\kappa^2M^4$ ($\equiv V_0$) dominates the universe. 
With SUSY thus broken, there are
radiative corrections from the $\phi$-$\bar{\phi}$ supermultiplets
that provide logarithmic corrections to the potential which drives inflation.
In one loop approximation \cite{hybrid,coleman}, 
\begin{eqnarray}\label{scalarpot}
V\approx V_0\bigg[1+\frac{\kappa^2N}{32\pi^2}\bigg(
4{\rm ln}\frac{\kappa|S|}{\Lambda}+(z+1)^2{\rm ln}(1+z^{-1})
+(z-1)^2{\rm ln}(1-z^{-1})\bigg)\bigg],~ 
\end{eqnarray}
where $z=x^2=|S|^2/M^2$, $\Lambda$ denotes a renormalization mass scale and 
$N$ denotes the dimensionality of the $\phi$, $\bar{\phi}$ representations.  
From Eq.~(\ref{scalarpot}) the quadrupole anisotropy is 
found to be~\cite{hybrid} 
\begin{eqnarray}\label{T}
\bigg(\frac{\delta T}{T}\bigg)_Q\approx \frac{8\pi}{\sqrt{N}}
\bigg(\frac{N_Q}{45}\bigg)^{1/2}\bigg(\frac{M}{M_{\rm Planck}}\bigg)^2
x_Q^{-1}y_Q^{-1}f(x_Q^2)^{-1} ~.  
\end{eqnarray}
The subscript $Q$ is there to emphasize the epoch of horizon crossing, 
$y_Q\approx x_Q(1-7/12x_Q^2+\cdots)$, $f(x_Q^2)^{-1}\approx 1/x_Q^2$, 
for $S_Q$ sufficiently larger than $M$, 
and $N_Q\approx 50-60$ denotes the e-foldings needed to resolve the horizon 
and flatness problems.  
From the expression for $\delta T/T$ in Eq.~(\ref{T}) and comparison with the 
COBE result $(\delta T/T)_Q\approx 6.6\times 10^{-6}$~\cite{cobe}, 
it follows that the gauge symmetry breaking scale $M$ is 
close to $10^{16}$ GeV.  Note that $M$ is associated in our $SO(10)$ example 
with the breaking scale of $H$ (in particular the $B-L$ breaking scale), 
which need not exactly coincide with the SUSY GUT scale.  
We will be more specific about $M$ later.     

The relative flatness of the potential ensures that the primordial density
fluctuations are essentially scale invariant. 
Thus, the scalar spectral index $n$ is
0.98 for the simplest example based on $W$ in Eq.~(\ref{simplepot}).

Several comments are in order:

\noindent $\bullet$ The 50-60 e-foldings required to solve the horizon and
flatness problems occur when the inflaton field $S$ is relatively close
(to within a factor of order 1-10) to the GUT scale.
Thus, Planck scale corrections can be safely ignored.

\noindent $\bullet$ For the case of minimal K${\rm\ddot{a}}$hler potential,
the SUGRA corrections do not affect the scenario at all,
which is a non-trivial result~\cite{hybrid}.  
More often than not, supersymmetric inflationary scenarios fail to work
in the presence of SUGRA corrections which tend to spoil the flatness
of the potential needed to realize inflation.

\noindent $\bullet$ Turning to the subgroup $H$ of $SO(10)$,
one needs to take into account the fact that the spontaneous breaking of $H$
produces magnetic monopoles that carry two quanta of 
Dirac magnetic charge~\cite{magg}.  
An overproduction of these monopoles at or near
the end of inflation is easily avoided, say by introducing an additional
(non-renormalizable) term $S(\phi\bar{\phi})^2$ in $W$,
which is permitted by the $U(1)_R$ symmetry.
The presence of this term ensures the absence of monopoles as explained
in Ref.~\cite{khalil}. Note that the monopole problem is also
avoided by choosing a different subgroup of $SO(10)$.

\noindent $\bullet$ At the end of inflation the scalar fields
$\phi$, $\bar{\phi}$, and $S$ oscillate about their respective minima.
Since the $\phi$, $\bar{\phi}$ belong respectively 
to the ${\bf (\overline{4},1,2)}$ 
and ${\bf (4,1,2)}$ of $SU(4)_c\times SU(2)_L\times SU(2)_R$, 
they decay exclusively into right handed neutrinos 
via the superpotential couplings, 
\begin{eqnarray} \label{nu}
W=\frac{\gamma_{i}}{M_P}\bar{\phi}\bar{\phi}F^c_iF^c_i ~,
\end{eqnarray} 
where the matter superfields $F^c_i$ belong to 
the ${\bf (\overline{4},1,2)}$ representation of $H$, and 
$M_P\equiv M_{\rm Planck}/\sqrt{8\pi}=2.44\times 10^{18}$ GeV denotes 
the reduced Planck mass, and $\gamma_{i}$ are dimensionless coefficients.  

\section{Inflationary Solution and Brane Gravity}

We consider 5D space-time ($x^\mu, y$),
$\mu=0,1,2,3$, where the fifth dimension is compactified on an $S^1/Z_2$
orbifold.  
The results in $S^2/Z_2$ could be readily applied
to the $S^1/(Z_2\times Z_2')$ case.  
The action is given by
\begin{eqnarray} \label{action}
S=\int d^4x\int_{-y_c}^{y_c}dy\sqrt{|g_5|}\bigg[\frac{M_5^3}{2}R_5
+\frac{\delta(y)}{\sqrt{g_{55}}}\bigg(\frac{M_4^2}{2}\bar{R}_4-\Lambda_1\bigg)
-\frac{\delta(y-y_c)}{\sqrt{g_{55}}}\Lambda_2\bigg],~
\end{eqnarray}
where $R_5$ ($\bar{R}_4$) is the 5 dimensional (4 dimensional) 
Einstein-Hilbert term\footnote{The importance of the 
brane-localized 4D Einstein-Hilbert term, especially for generating 4D gravity
in a higher dimensional non-compact flat space was first noted
in Ref.~\cite{braneR}.},     
and $\Lambda_1$, $\Lambda_2$ are
the brane cosmological constants. 
Note that the bulk cosmological constant is not introduced in the action.  
$M_5$ and $M_4$ are mass parameters.
The cosmological constants on the branes could be interpreted
the vacuum expectation values of some scalar potentials
from the particle physics sector.
The brane curvature scalar (Ricci scalar) $\bar{R}_4(\bar{g}_{\mu\nu})$
is defined with the induced metric of the bulk metric,
$\bar{g}_{\mu\nu}(x)\equiv g_{\mu\nu}(x,y=0)$ 
($\mu,\nu=0,1,2,3$).  
For an inflationary solution, we take the metric ansatz,
\begin{eqnarray} \label{metric}
ds^2=\beta^2(y)(-dt^2+e^{2H_0t}d\vec{x}^2)+dy^2 ~,
\end{eqnarray}
where $H_0$ could be interpreted as the 4 dimensional Hubble constant.
The non-vanishing components $(\mu,\mu)$ and $(5,5)$
of the 5 dimensional Einstein equation
derived from (\ref{action}) gives 
\begin{eqnarray} \label{eom}
&&3\bigg[\bigg(\frac{\beta'}{\beta}\bigg)^2+\bigg(\frac{\beta''}{\beta}\bigg)
-\bigg(\frac{H_0}{\beta}\bigg)^2
-\delta(y)\frac{M_4^2}{M_5^3}\bigg(\frac{H_0}{\beta}\bigg)^2\bigg] \\
&&~~~~~~~~~~~=-\delta(y)\frac{\Lambda_1}{M_5^3}
-\delta(y-y_c)\frac{\Lambda_2}{M_5^3} ~,  \nonumber \\
&&6\bigg[\bigg(\frac{\beta'}{\beta}\bigg)^2-\bigg(\frac{H_0}{\beta}\bigg)^2
\bigg]=0 ~, \label{eom2}
\end{eqnarray}
where primes denote derivatives with respect to $y$.
The last term in the left hand side in Eq.~(\ref{eom})
arises from the brane scalar curvature term, and vanishes when $H_0=0$.

The solutions to the equations Eq.~(\ref{eom}) and (\ref{eom2}) is given by
\begin{eqnarray} \label{sol1}
\beta(y)&=&\pm H_0|y|+c 
\end{eqnarray}
where $c$ ($\approx 1$) is an integration constant.  
The introduction of the brane scalar curvature term $\bar{R}_4$
does not affect the bulk solutions (\ref{sol1}),
but it modifies the boundary conditions at $y=0$ and $y=y_c$,  
\begin{eqnarray} 
&&\pm\frac{H_0}{c}-\frac{1}{2}\frac{M_4^2}{M_5^3}~\frac{H_0^2}{c^2}
=-\frac{\Lambda_1}{6M_5^3} ~, \label{bdy1'} \\
&&\frac{\pm H_0}{c\pm H_0y_c}=\frac{\Lambda_2}{6M_5^3} ~.  \label{bdy2'}
\end{eqnarray}    
We note that $\Lambda_1$ and $\Lambda_2$ are related
to the 4 dimensional Hubble constant $H_0$.
While their non-zero values are responsible for the 3-space
inflation, vanishing brane cosmological constants guarantee 
a 4 dimensional flat space-time.
When $\Lambda_1=0$, $\Lambda_2$ must be also zero.
Hence it is natural that the scalar field which controls inflation
is introduced in the bulk.~\footnote{
Since SUSY is broken at low energies, the minima of the inflaton potentials 
on both branes should be fine-tuned to zero.~\cite{selftun}
}

For $\Lambda_2>0$ and $c>H_0y_c$ ($\sim H_0/M_{\rm GUT}$),
`$+$' is chosen in Eq.~(\ref{sol1}).  
From Eqs.~(\ref{bdy1'})--(\ref{bdy2'}), 
we also note that the brane cosmological
constants $\Lambda_1$ and $\Lambda_2$ should have opposite signs
in the absence of the brane curvature scalar contribution at $y=0$.
However, a suitably large value of $M_4/M_5$ can even make
the sign of $\Lambda_1$ positive.
Since the introduction of the brane curvature term does not conflict with
any symmetry that may be present,
there is no reason why such a term with a parameter $M_4$ that is large
compared to $M_5$ is not allowed~\cite{braneR}.
Thus, $\Lambda_1$ and $\Lambda_2$ could both be positive
and this fact will be exploited for implementing the inflationary scenario.

The condition for a positive brane cosmological constant
on B1 is found from (\ref{bdy1'}) to be $(H_0/c)(M_4^2/M_5^3)>2$.
For $\kappa\sim 10^{-3}$, say, and $c\sim 1$, we have
$H_0\sim 10^{11}$ GeV and $M_5\sim 10^{16}$ GeV (so that $M_4\sim M_P$).
Thus, there exists a hierarchy of order $10^{2}$
between the 5D bulk scale $M_5$ and
the four dimensional brane mass scale $M_4$.
One could construct a simple model to explain it~\cite{original}.

Our main task is to embed the 4D supersymmetric inflationary scenario
in 5D space-time, 
employing the framework and solutions discussed above.
In order to extend the setup to 5D SUGRA, a gravitino $\psi_M$
and a vector field $B_M$ should be appended
to the graviton (f${\rm\ddot{u}}$nfbein) $e_{M}^{m}$.
Through orbifolding, only $N=1$ SUSY is preserved on the branes.    
The brane-localized Einstein-Hilbert term in Eq.~(\ref{action}) 
is still allowed, but should 
be accompanied by a brane gravitino kinetic term as well as other terms,
which is clear in off-shell SUGRA formalism \cite{kyae}.  
In a higher dimensional supersymmetric theory, a F-term scalar potential is
allowed only on the 4 dimensional fixed points 
which preserve $N=1$ SUSY.  
We require a formalism in which inflation and
the Hubble constant $H_0$ are controlled
only by the brane cosmological constants, such that
during inflation the positive vacuum energy slowly decreases, and
the minimum of the scalar potential corresponds to a flat 4D space-time.  
The boundary conditions (\ref{bdy1'}) and (\ref{bdy2'}) meet 
these requirements
in the presence of the additional brane scalar curvature term at $y=0$.

\section{5D $SO(10)$ Model on $S^1/(Z_2\times Z_2')$}

Let us consider the 4D $SU(4)_c\times SU(2)_L\times SU(2)_R (\equiv H)$
supersymmetric inflationary model \cite{khalil}.
An effective 4D theory with the gauge group $H$ is readily obtained
from a 5D $SO(10)$ gauge theory if the fifth dimension is compactified
on the orbifold $S^1/(Z_2\times Z_2')$ \cite{dermisek}, where 
$Z_2$ reflects $y\rightarrow -y$, and $Z_2'$ reflects $y'\rightarrow -y'$  
with $y'=y+y_c/2$.
There are two independent orbifold fixed points (branes)
at $y=0$ and $y=y_c/2$, with $N=1$ SUSYs and gauge symmetries 
$H$ and $SO(10)$ respectively~\cite{dermisek}.  
The $SO(10)$ gauge multiplet $(A_M,\lambda^1,\lambda^2,\Phi)$ 
decomposes under $H$ as
\begin{eqnarray}
V_{\bf 45}&\longrightarrow& V_{({\bf 15},{\bf 1},{\bf 1})}+
V_{({\bf 1},{\bf 3},{\bf 1})}+V_{({\bf 1},{\bf 1},{\bf 3})}
+V_{({\bf 6},{\bf 2},{\bf 2})} \\
&&+\Sigma_{({\bf 15},{\bf 1},{\bf 1})}
+\Sigma_{({\bf 1},{\bf 3},{\bf 1})}+\Sigma_{({\bf 1},{\bf 1},{\bf 3})}
+\Sigma_{({\bf 6},{\bf 2},{\bf 2})} ~,  \nonumber
\end{eqnarray}
where $V$ and $\Sigma$ denote the vector multiplet $(A_\mu,\lambda^1)$ and
the chiral multiplet $((\Phi+iA_5)/\sqrt{2},\lambda^2)$ respectively,
and their $(Z_2,Z_2')$ parity assignments and KK masses
are shown in Table I.
\vskip 0.6cm
\begin{center}
\begin{tabular}{|c||c|c|c|c|} \hline
Vector &$V_{({\bf 15},{\bf 1},{\bf 1})}$ & $V_{({\bf 1},{\bf 3},{\bf 1})}$ &
$V_{({\bf 1},{\bf 1},{\bf 3})}$ &$V_{({\bf 6},{\bf 2},{\bf 2})}$ \\
\hline \hline
$(Z_2,Z_2')$ &(+,+)$$ &$(+,+)$ &$(+,+)$ &$(-,+)$ \\
Masses & $2n\pi/y_c$ & $2n\pi/y_c$ & $2n\pi/y_c$ & $(2n+1)\pi/y_c$ \\
\hline \hline
Chiral & $\Sigma_{({\bf 15},{\bf 1},{\bf 1})}$ &
$\Sigma_{({\bf 1},{\bf 3},{\bf 1})}$ &
$\Sigma_{({\bf 1},{\bf 1},{\bf 3})}$ &$\Sigma_{({\bf 6},{\bf 2},{\bf 2})}$
\\ \hline \hline
$(Z_2,Z_2')$ & $(-,-)$ &$(-,-)$ &$(-,-)$ &$(+,-)$  \\
Masses & $(2n+2)\pi/y_c$ & $(2n+2)\pi/y_c$ & $(2n+2)\pi/y_c$ & $(2n+1)\pi/y_c$
\\
\hline
\end{tabular}
\end{center}
{\bf Table I.~}($Z_2,Z_2'$) parity assignments and
Kaluza-Klein masses $(n=0,1,2,\cdots)$ for the vector multiplet 
in $N=2$ SUSY $SO(10)$.
\vskip 0.6cm

The parities of the chiral multiplets $\Sigma$'s are opposite to those of
the vector multiplets $V$'s in Table I and
hence, $N=2$ SUSY explicitly breaks to $N=1$
below the compactification scale $\pi/y_c$.
As shown in Table I, only the vector multiplets,
$V_{({\bf 15},{\bf 1},{\bf 1})}$, $V_{({\bf 1},{\bf 3},{\bf 1})}$, and
$V_{({\bf 1},{\bf 1},{\bf 3})}$ contain massless modes, which
means that the low energy effective 4D theory reduces to $N=1$
supersymmetric $SU(4)_c\times SU(2)_L\times SU(2)_R$.
The parity assignments in Table I also show that the wave function
of the vector multiplet $V_{({\bf 6},{\bf 2},{\bf 2})}$ vanishes
at the brane located at $y=0$ (B1) because it is assigned an odd parity
under $Z_2$, while the wave functions of all the vector multiplets should
be the same at the $y=y_c/2$ brane (B2).
Therefore, while the gauge symmetry at B2 is $SO(10)$,
only $SU(4)_c\times SU(2)_L\times SU(2)_R$ is preserved at B1 \cite{hebecker}.

The inflationary solution requires positive vacuum energies on both branes 
B1 and B2.  While the scalar potential in Eq.~(\ref{scalarpot}) 
would be suitable for B1, an appropriate scalar potential on B2 is 
also required.  
Since the boundary conditions in Eq.~(\ref{bdy1'}) and (\ref{bdy2'})
require $\Lambda_1$ and $\Lambda_2$ to simultaneously vanish, 
it is natural to require $S$ to be a bulk field.    
Then, the VEVs of $S$ on the two branes can be adjusted
such that the boundary conditions are satisfied.
As an example, consider the following superpotential on B2,
\begin{eqnarray} \label{b1superpot}
W_{B2}=\kappa_1S(Z\overline{Z}-M_1^2) ~,
\end{eqnarray}
where $Z$ and $\overline{Z}$ are $SO(10)$ singlet superfields on the B1 brane 
with opposite $U(1)_R$ charges.  

\section{Leptogenesis} 

After inflation is over, the oscillating system consists of the complex scalar 
fields $\Phi=(\delta\bar{\phi}+\delta\phi)$, 
where $\delta\bar{\phi}=\bar{\phi}-M$ ($\delta \phi=\phi-M$), and $S$,  
both with masses equal to $m_{\rm infl}=\sqrt{2}\kappa M$.  
Through the superpotential couplings in Eq.~(\ref{nu}), 
these fields decay into a pair of right handed neutrinos and sneutrinos 
respectively, with an approximate decay width \cite{khalil}
\begin{eqnarray}\label{decay}
\Gamma\sim \frac{m_{\rm infl}}{8\pi}\bigg(\frac{M_i}{M}\bigg)^2 ~, 
\end{eqnarray}
where $M_i$ denotes the mass of the heaviest right handed neutrino 
with $2M_i<m_{\rm infl}$, so that the inflaton decay is possible.  
Assuming an MSSM spectrum below the GUT scale, the reheat temperature 
is given by \cite{hybrid2}
\begin{eqnarray} \label{temp}
T_r\approx \frac{1}{3}\sqrt{\Gamma M_P}
\approx\frac{1}{12}\bigg(\frac{55}{N_Q}\bigg)^{1/4}\sqrt{y_Q}M_i ~.  
\end{eqnarray} 
For $y_Q\sim$ unity (see below), and $T_r\lapproxeq 10^{9.5}$ GeV 
from the gravitino constraint~\cite{gravitino}, 
we require $M_i\lapproxeq 10^{10}-10^{10.5}$ GeV. 

In order to decide on which $M_i$ is involved in the decay~\cite{pati}, 
let us start with atmospheric neutrino ($\nu_\mu-\nu_\tau$) oscillations and 
assume that the light neutrinos exhibit an hierarchical mass pattern  
with $m_3>>m_2>>m_1$.   
Then $\sqrt{\Delta m^2_{\rm atm}}\approx m_3\approx m_{D3}^2/M_3$, 
where $m_{D3}$ ($=m_t(M)$) 
denotes the third family Dirac mass which equals the asymptotic top quark 
mass due to $SU(4)_c$.  We also assume a mass hierarchy in the right handed 
sector, $M_3>>M_2>>M_1$.  
The mass $M_3$ arises from the superpotential coupling Eq.~(\ref{nu}) 
and is given by 
$M_3= 2\gamma_{3}M^2/M_P\sim 10^{14}~~{\rm GeV}$,  
for $M\sim 10^{16}$ GeV and $\gamma_{3}\sim$ unity.  This value of $M_3$ is in 
the right ball park to generate an $m_3\sim \frac{1}{20}$ eV 
($\sim \sqrt{\Delta m_{\rm atm}^2}$), 
with $m_t(M)\sim 110$ GeV \cite{hybrid2}.  
It follows from (\ref{temp}) that $M_i$ in (\ref{decay}) cannot be identified 
with the third family right handed neutrino mass $M_3$.  
It should also not correspond to the second family neutrino mass $M_2$ 
if we make the plausible assumption that the second generation Dirac mass 
should lie in the few GeV scale.  
The large mixing angle MSW solution of the solar neutrino 
problem requires that 
$\sqrt{\Delta m_{\rm solar}^2}\approx m_2\sim {\rm GeV}^2/M_2 
\sim \frac{1}{160}~{\rm eV}$, 
so that $M_2\gapproxeq 10^{11}-10^{12}$ GeV.
Thus, we are led to conclude~\cite{pati} that 
the inflaton decays into the lightest (first family) right handed neutrino  
with mass 
\begin{eqnarray}\label{M1} 
M_1\sim 10^{10}-10^{10.5} ~{\rm GeV} ~, 
\end{eqnarray} 
such that $2M_1<m_{\rm infl}$.  

The constraint $2M_2> m_{\rm infl}$ yields 
$y_Q\lapproxeq 3.34 \gamma_{2}$, 
where $M_2=2\gamma_{2}M^2/M_P$.  We will not provide here a comprehensive 
analysis of the allowed parameter space but will be content to present 
a specific example, namely 
\begin{eqnarray}\label{value}
M\approx 8\times 10^{15}~{\rm GeV}~,~~\kappa\approx 10^{-3}~,~~
m_{\rm infl}\sim 10^{13}~{\rm GeV}~(\sim M_2)~, 
\end{eqnarray}
with $y_Q\approx 0.4$ (corresponding to $x_Q$ near unity, so that the inflaton 
$S$ is quite close to $M$ during the last 50--60 e-foldings).   

Note that typically $\kappa$ is of order 
$10^{-2}$-- few $\times 10^{-4}$~\cite{khalil}, 
so that the vacuum energy density during inflation is 
$\sim 10^{-4}-10^{-8}~M_{\rm GUT}^4$.  
Thus, in this class of models 
the tensor to scalar ratio $r$ is highly suppressed, $r \lapproxeq 10^{-4}$.    
%
%
With $\kappa\sim{\rm few}\times 10^{-4}$ ($10^{-3}$), 
the scalar spectral index $n\approx 0.99$ ($0.98$).   

The decay of the (lightest) right handed neutrinos generates 
a lepton asymmetry which is given by \cite{lasymm} 
\begin{eqnarray} \label{lasymm}
\frac{n_L}{s}\approx \frac{10}{16\pi}\bigg(\frac{T_r}{m_{\rm infl}}\bigg)
\bigg(\frac{M_1}{M_2}\bigg)
\frac{c_{\theta}^2s_{\theta}^2~{\rm sin}2\delta~(m_{D2}^2-m_{D1}^2)^2}
{|\langle h\rangle|^2(m_{D2}^2s_{\theta}^2+m_{D1}^2c_{\theta}^2)} ~, 
\end{eqnarray} 
where the VEV $|\langle h\rangle|\approx 174$ GeV 
(for large ${\rm tan}\beta$), 
$m_{D1,2}$ are the neutrino Dirac masses (in a basis in which they are 
diagonal and positive), and $c_\theta\equiv {\rm cos}\theta$, 
$s_\theta\equiv {\rm sin}\theta$, with $\theta$ and $\delta$ being the rotation 
angle and phase which diagonalize the Majorana mass matrix of 
the right handed neutrinos.  
Assuming $c_\theta$ and $s_\theta$ of comparable magnitude, 
taking $m_{D2}>>m_{D1}$, and using (\ref{M1}) and (\ref{value}), 
Eq.~(\ref{lasymm}) reduces to   
\begin{eqnarray}
\frac{n_L}{s}\approx 10^{-8.5}c_\theta^2{\rm sin}2\delta 
\bigg[\frac{T_r}{10^{9.5}{\rm GeV}}\cdot
\frac{M_1}{2\cdot 10^{10.5}{\rm GeV}}\cdot
\frac{10^{13}{\rm GeV}}{M_2}\cdot
\frac{m_{D2}^2}{100{\rm GeV^2}}\bigg],~
\end{eqnarray}
which can be in the correct ball park to account for the observed baryon 
asymmetry $n_B/s$ ($\approx -28/79 ~n_L/s$).  

\section{Conclusion}

We have proposed a realistic model, 
which nicely blends together four particularly 
attractive ideas, namely supersymmetric grand unification, extra dimension,  
inflation and leptogenesis.  
To accomodate a 4D F-term inflationary model in 5D,
a brane gravity term is necessary.  
The doublet-triplet problem is circumvented by utilizing orbifold
breaking of $SO(10)$, which may also help in suppressing dimension five
proton decay.
Concerning inflation, 
the scalar spectral index $n$ lies very close to unity 
($\approx 0.98-0.99$), and 
the tensor to scalar ratio $r$ is highly suppressed 
($\lapproxeq 10^{-4}$).  
%
%
Finally, the inflaton decay produces heavy right handed Majorana neutrinos 
(in our case the lightest one), whose subsequent out of equilibrium decay 
leads to the baryon asymmetry via leptogenesis.  

\section*{Acknowledgments} 
The work is partially supported
by DOE under contract number DE-FG02-91ER40626.

\end{document}